\documentclass[aps,pre,reprint,groupedaddress]{revtex4-1}
\usepackage[utf8]{inputenc}
\usepackage{amssymb}
\usepackage{graphicx,amsmath}
\usepackage[colorlinks=true,linkcolor=black, citecolor=blue, urlcolor=blue]{hyperref}
\usepackage{longtable}
\usepackage{appendix}
\usepackage{booktabs}

\usepackage{tabularx}
\usepackage{ltablex}
\usepackage{xcolor}

\usepackage{braket}
\usepackage{xr}

\usepackage{mathtools}

\DeclarePairedDelimiter\floor{\lfloor}{\rfloor}

\newcommand{\nocontentsline}[3]{}
\newcommand{\tocless}[2]{\bgroup\let\addcontentsline=\nocontentsline#1{#2}\egroup}

\begin{document}


\title{Identification of milestone papers through time-balanced network centrality}


\author{Manuel Sebastian Mariani}
\email{manuel.mariani@unifr.ch}
\affiliation{Department of Physics, University of Fribourg, 1700 Fribourg, Switzerland}
\author{Mat{\'u}{\v{s}} Medo}
\affiliation{Department of Physics, University of Fribourg, 1700 Fribourg, Switzerland}
\author{Yi-Cheng Zhang}
\affiliation{Department of Physics, University of Fribourg, 1700 Fribourg, Switzerland}
\affiliation{Chongqing Institute of Green and Intelligent Technology, Chinese Academy of Sciences, Chongqing 400714, P.R. China}


\begin{abstract}
Citations between scientific papers and related bibliometric indices, such as the $h$-index for authors and the impact factor for journals,
are being increasingly used -- often in controversial ways -- as quantitative tools for research evaluation.
Yet, a fundamental research question remains still open: to which extent do quantitative
metrics capture the significance of scientific works? 
We analyze the network of citations among the $449,935$ papers published
by the American Physical Society (APS) journals between $1893$ and $2009$, and focus on the comparison of metrics built on the citation
count with network-based metrics. 
We contrast five article-level metrics with respect to the rankings that they assign to a set of
fundamental papers, called Milestone Letters, carefully selected by the APS editors for “making long-lived contributions
to physics, either by announcing significant discoveries, or by initiating new areas of research”.
A new metric, which combines
PageRank centrality with the explicit requirement that paper score is not biased by paper age,  is the best-performing metric overall in identifying
the Milestone Letters. 
The lack of time bias in the new metric makes it also possible to use it to compare
papers of different age on the same scale. We find that network-based metrics identify the Milestone Letters better than metrics
based on the citation count, which suggests that the structure of the citation network contains information that can be used to improve
the ranking of scientific publications. The methods and results presented here are relevant for all evolving systems where
network centrality metrics are applied, for example the World Wide Web and online social networks.
An interactive Web platform where it is possible to view the ranking of the APS papers by rescaled PageRank
is available at the address \url{http://www.sciencenow.info}.
\end{abstract}


\maketitle


\section{Introduction}

The notion of quantitative evaluation of scientific impact
builds on the basic idea that the scientific merits of papers \cite{narin1976evaluative, radicchi2008universality}, 
scholars \cite{hirsch2005index, egghe2006theory},
journals \cite{pinski1976citation, liebowitz1984assessing,bollen2006journal}, 
universities \cite{kinney2007national,molinari2008new} and
countries \cite{king2004scientific, cimini2014scientific}
can be gauged by metrics based on the received citations.
The respective field, referred to as bibliometrics or scientometrics,
is undergoing a rapid growth \cite{van2010metrics} fueled 
by the increasing availability of massive citation datasets
collected by both academic journals and online platforms, such as Google Scholar and Web of Science.
The possible benefits, drawbacks
and long-term effects of the use
of bibliometric indices are being highly debated by scholars from diverse
fields \cite{van2005fatal, weingart2005impact, lawrence2008lost, hicks2015bibliometrics, werner2015focus}.

Although some effort has been devoted to contrast different metrics with respect to 
their ability to single out seminal papers \cite{dunaiski2012comparing,yao2014ranking, zhou2015ranking,dunaiski2016evaluating}, 
differences among the adopted benchmarking procedures and 
diverse conclusions of the mentioned references 
leave a fundamental question still open:
which metric of scientific impact best agrees with expert-based perception of significance?
In agreement with ref. \cite{wasserman2015cross}, the significance of a scientific work is intended here as its 
enduring importance within the scientific community.

To address this question, we focus on
a list of $87$ physics papers of outstanding significance -- called Milestone Letters -- recently made available by
the American Physical Society (APS)  [\url{http://journals.aps.org/prl/50years/milestones},
accessed 25-11-2015].
According to the APS editors' description, the Milestone Letters
``have made long-lived contributions to physics, either by announcing significant discoveries, or by initiating new areas of research''.
These articles have been carefully selected by the editors of the APS,
and the choices are motivated in detail in the webpage;
the fact that a large fraction of them led to Nobel Prize for some of their authors is an indication of the
exceptional level of the selected works.

In this work,
we analyze the network of citations between the $N=449,935$ papers published in APS journals from $1893$ until $2009$
to compare five article-level metrics with respect to the ranking position they assign to
the Milestone Letters.
A reliable expert-based evaluation of the significance (intended as enduring importance, as in ref. \cite{wasserman2015cross}) 
of a paper necessarily requires a time lag
between the paper's publication date and the expert's judgment. For example, there is a time interval of $14$ years
between the most recent Milestone Letter 
(from $2001$) and the year at which the list of Milestone Letters was released ($2015$).
However, 
we show that a well-designed quantitative metric offers us the opportunity to
detect potentially significant papers relatively short after their publication -- an aspect often neglected in the evaluation of bibliometric indicators.
To show this, we study how the ability of the different metrics to identify the Milestone Letters changes with paper age.

A plethora of quantitative metrics exist and could be studied in principle.
Our focus here is narrowed to metrics that rely on a diffusion process on the underlying network of citations between papers
and their comparison with simple citation count.
The five metrics considered in this work are thus:
the citation count, 
PageRank \cite{brin1998anatomy}, 
CiteRank \cite{walker2007ranking}, rescaled citation count \cite{newman2009first},
and novel rescaled PageRank.
PageRank is a classical network centrality metric which combines a random walk along network links with
a random teleportation process. 
The metric
has been applied to a broad range of real-world problems
(see refs. \cite{franceschet2011pagerank,gleich2015pagerank,ermann2015google} for a review),
including ranking academic papers \cite{chen2007finding, yao2014ranking}, journals \cite{bollen2006journal,gonzalez2010new} 
and authors \cite{radicchi2009diffusion, yan2009applying, nykl2014pagerank} (see ref. \cite{waltman2014pagerank} for a review of
the applications of PageRank-related methods
to bibliometric analysis). 

To overcome the well-known PageRank's bias toward old nodes in citation data (detailedly studied in refs.
\cite{chen2007finding,mariani2015ranking}),
the CiteRank algorithm introduces exponential penalization of old nodes, resulting in a node score that
well captures the future citation count increase of the papers and, for this reason, can be considered as a reasonable proxy for
network traffic, as shown by \cite{walker2007ranking}.
However, we show below that CiteRank score does not allow one to fairly compare papers of different age.
Rescaled citation count and rescaled PageRank are derived from citation count and PageRank score,
respectively, by explicitly requiring that paper score is not biased by age -- 
the adopted rescaling procedure is conceptually close to the methods recently developed in refs.
\cite{radicchi2008universality,newman2009first,radicchi2011rescaling,newman2014prediction,radicchi2012testing,radicchi2012reverse,crespo2012citation,kaur2015quality}
to suppress biases by age and field in
the evaluation of academic agents. 
We find that the rankings produced by the rescaled scores are indeed consistent with the hypothesis that the rankings are not biased by age.

We find that PageRank can compete and even outperform
rescaled PageRank in identifying \emph{old} milestone
papers,
but completely fails to identify recent milestone papers due to its temporal bias.
CiteRank can compete and even outperform
rescaled PageRank in identifying \emph{recent} milestone
papers,
but markedly underperforms in identifying old milestone papers due to its built-in exponential penalization for older papers.
Indicators based on simple citation count are outperformed by rescaled PageRank for papers of every age.
This leads us to the conclusion that rescaled PageRank is the best-performing metric overall.
With respect to previous works \cite{chen2007finding,dunaiski2012comparing,fiala2012time, dunaiski2016evaluating}
that claimed the superiority of network-based metrics in identifying important
papers, 
our results clarify the essential role of paper age in determining the metrics'
performance:
rescaled PageRank excels and PageRank performs poorly in identifying MLs short after their publication, 
and the performance of the two methods becomes comparable only $15$ years after the MLs are published.
Qualitatively similar results are found for an alternative list of APS outstanding papers 
which only includes works that have led to Nobel prize for some of the authors
(the list is provided in the Table S1).

Our results indicate that network centrality
and time-balance are two essential ingredients -- though neglected by popular bibliometric indicators such as the
$h$-index for scholars \cite{hirsch2005index} and impact factor for journals \cite{garfield1972citation} -- for an effective detection of significant papers.
This sets a new benchmark for article-level metrics and quantitatively support the paradigm that considering the whole network instead of 
simple citation count can bring substantial benefits to the ranking of academic agents.
In a broader context, our results show that
a direct rescaling of PageRank scores is an effective method 
to solve the PageRank's well-known bias
against recent network nodes. We emphasize that while scientific papers are the focus of this work,
the addressed research question is general and can emerge when estimating the importance of any creative work -- such as movies 
\cite{spitz2014measuring, wasserman2015cross} -- 
for which quantitative impact metrics and expert-based significance assessments are simultaneously available.
The potential broader applications and possible limitations of our results are discussed in the Discussion section.



\section{Metrics}
\label{sec:metrics}

We consider five article-level metrics: citation count $c$, PageRank score $p$, CiteRank score $T$,
rescaled PageRank score $R(p)$, and rescaled citation count $R(c)$.

\subsection{Citation count}

We denote by $\mathbf{A}$ the network's adjacency matrix ($A_{ij}$ is one if node $j$ points to node $i$ and zero otherwise).
Citation count (referred to as indegree in network science literature \cite{newman2010networks})
is one of the simplest metrics to infer node centrality in a network, being simply defined as
$c_i=\sum_j A_{ij}$ for a node $i$.
Citation count is the building block of the majority of metrics for assessing the impact of single papers, authors, journals 
(for a review of citation-based impact indicators see ref. \cite{waltman2016review}).

\subsection{PageRank}

The PageRank score vector \cite{brin1998anatomy} can be defined as the stationary state of a process which combines
a random walk along the network links and random teleportation.
 In a directed monopartite network composed of $N$ nodes,
the vector of PageRank scores $\{p_{i}\}$ can be found as the stationary solution of the following set of recursive linear equations
\begin{equation}
 p_{i}^{(n+1)}=\alpha\,\sum_{j:k_j^{out}>0}A_{ij}\,\frac{p_{j}^{(n)}}{k^{out}_{j}}+\alpha\,\sum_{j:k_j^{out}=0}\frac{p_{j}^{(n)}}{N}+\frac{1-\alpha}{N},
\label{pr}
\end{equation}
where $k^{out}_{j}:=\sum_{l}A_{lj}$ 
is the outdegree of node $j$,
$\alpha$ is the teleportation parameter, and $n$ is the iteration number.
Eq. \eqref{pr} represents the master equation of a diffusion process on the network, which converges to a unique stationary state 
independently of the initial condition (see ref. \cite{berkhin2005survey} for the mathematical details).
The PageRank score
$p_{i}$ of node $i$ can be interpreted as the average fraction
of time spent on node $i$ by a random walker who
with probability $\alpha$ follows the network's links and with probability $1-\alpha$ teleports to a random node.
Throughout this paper, we set $\alpha=0.5$ which is the usual choice for citation data \cite{chen2007finding}.

\subsection{CiteRank}

To correct the PageRank's strong temporal bias in citation networks, the
CiteRank algorithm \cite{walker2007ranking} introduces ad-hoc penalization for older nodes.
The CiteRank score $T$ is defined similarly as PageRank; 
differently from PageRank, in CiteRank equations the teleportation probability decays exponentially with
paper age with a certain timescale $\tau$.
According to refs. \cite{walker2007ranking, maslov2008promise}, this choice of the
teleportation vector is intended to favor the recent nodes and thus lead to a score that
better represents papers' relevance for the current lines of research. 
Using the same notation as Eq. \eqref{pr},
the vector of CiteRank scores $\{T_{i}\}$ can be found as the stationary solution of the following set of recursive linear equations
\begin{equation}
\begin{split}
 T_{i}^{(n+1)}&=\alpha\,\sum_{j:k_j^{out}>0}A_{ji}\,\frac{T_{j}^{(n)}}{k^{out}_{j}}+\alpha\,\sum_{j:k_j^{out}=0}\frac{T_{j}^{(n)}}{N}\\
 &+(1-\alpha)\frac{\exp{(-(t-t_{i})/\tau)}}{\sum_{j=1}^{N}\exp{(-(t-t_{j})/\tau)}},
 \end{split}
 \label{cr}
\end{equation}
where we denote by $t_{i}$ the publication date of paper $i$ and $t$ the time at which the scores are computed.
Throughout this paper we set $\alpha=0.5$ and $\tau=2.6$ years, which are the parameters chosen in ref. \cite{walker2007ranking}.
The performance of the algorithm for other values of the parameter $\tau$ is discussed in the caption of Fig. \ref{fig:citerank}, in 
\ref{appendix_citerank}.
We show below that exponential penalization of older nodes is not effective in removing PageRank's
bias, and propose instead a rescaled PageRank score $R(p)$ whose average value and standard deviation
do not depend on paper age.

 \begin{figure*}[t]
 \centering
  \includegraphics[scale=0.7,  angle=0]{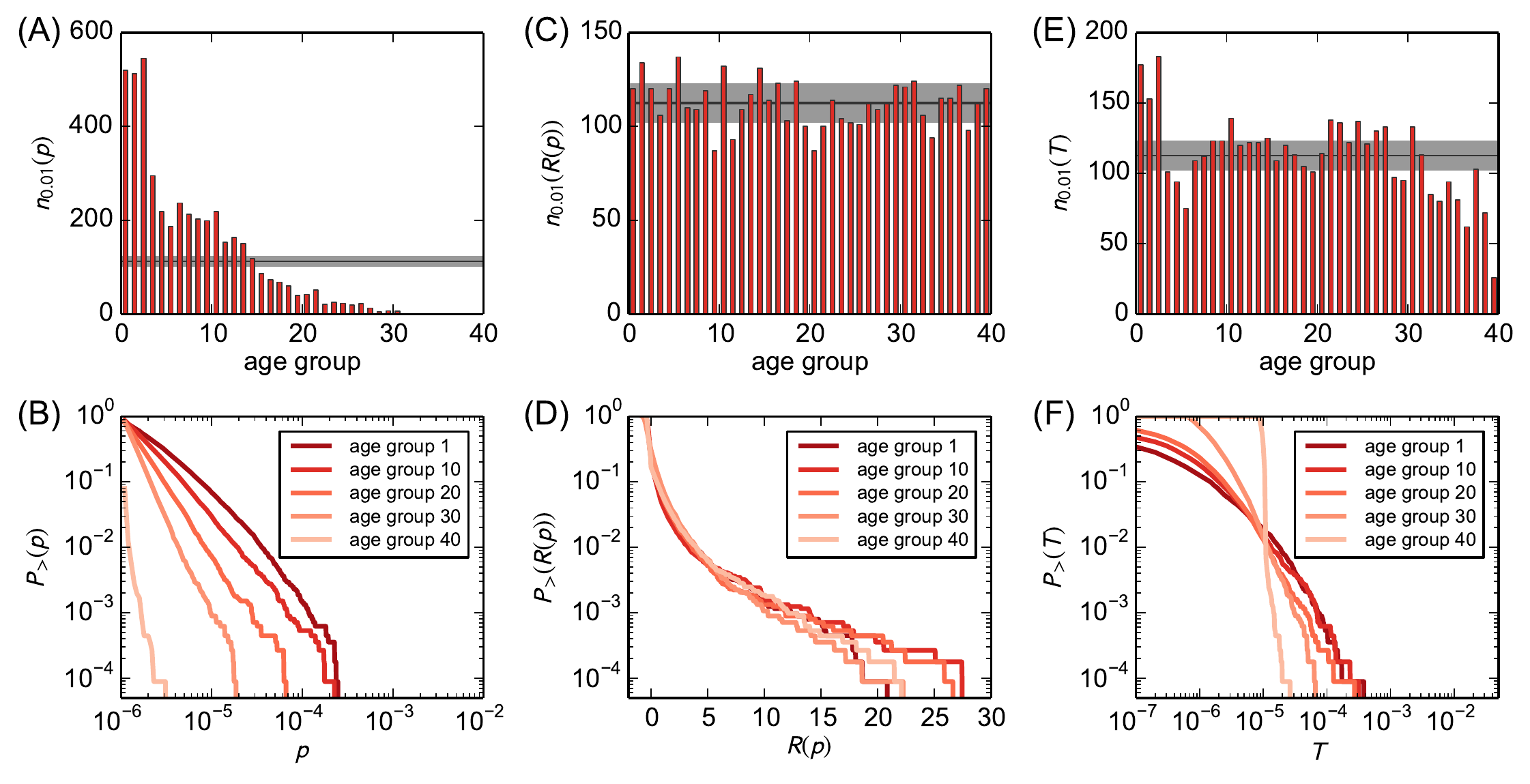}
 \caption{Time balance of the network-based metrics.
 Panels (A,C,E) show the histogram of the number of papers from each paper age group in the top-$1\%$
 of the ranking by PageRank score $p$, rescaled PageRank score $R(p)$ and CiteRank score $T$, respectively
 (age group $1$ and age group $40$ contain the oldest and most recent $N/40$ papers, respectively).
 The horizontal black line represents the unbiased value $n^{(0)}(0.01)=0.01\,N/40$; the gray-shaded area represents the interval
 $[n^{(0)}-\sigma_0,n^{(0)}+\sigma_0]$ with $\sigma_0$ given by Eq.~(\ref{sigma_0}).
 Panels (B,D,F) show the cumulative distributions 
 of PageRank score $p$, rescaled PageRank score $R(p)$ and CiteRank score $T$, respectively, for different age groups.
 \label{fig:one}}
\end{figure*}

\begin{figure*}[t]
\centering
 \includegraphics[scale=0.7,angle=0]{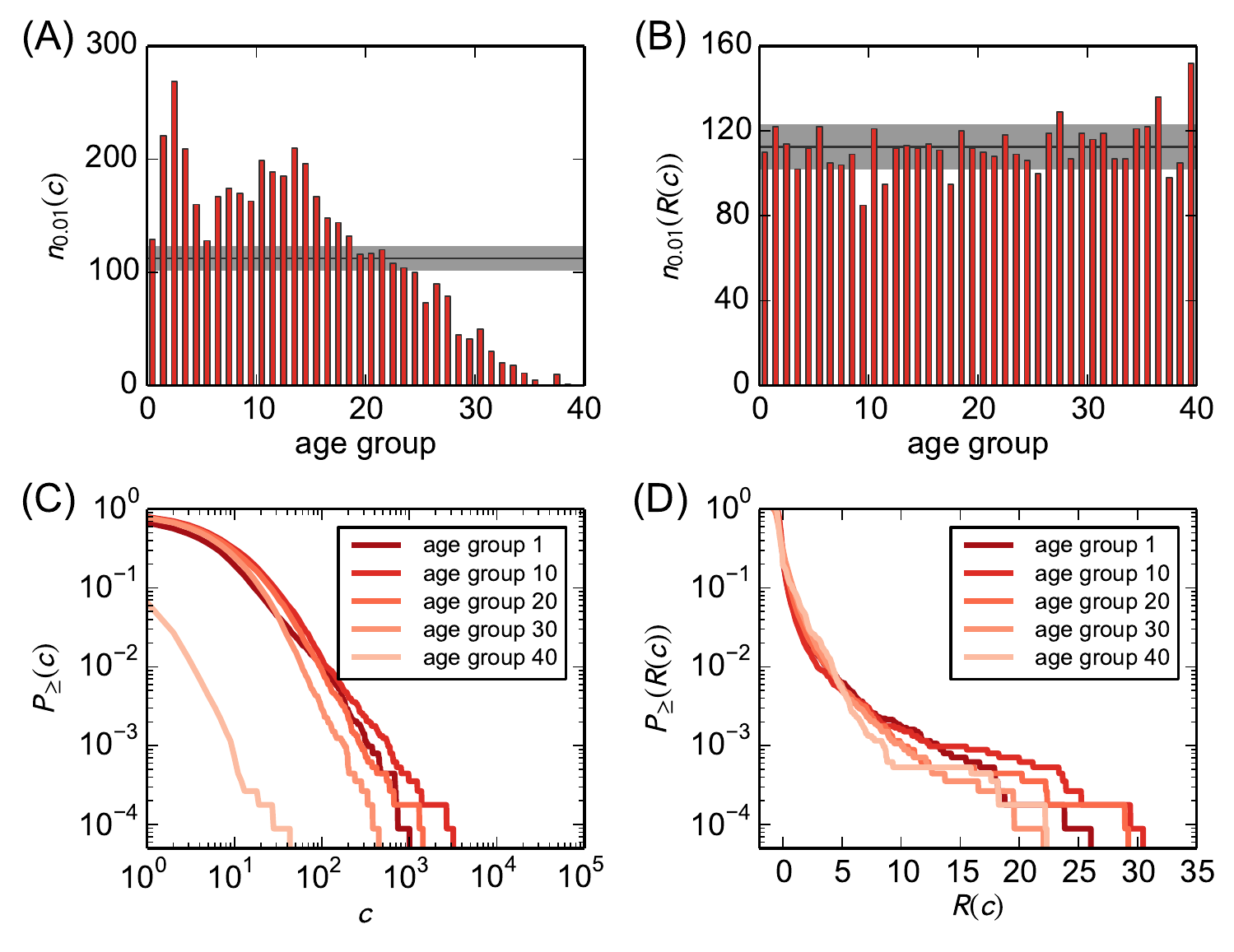}
\caption{Time balance of the citation-based metrics. Panels (A,B) show the histogram of the number of top-$1\%$ papers for each paper age group
 in the ranking by citation count $c$ and rescaled citation count $R(c)$, respectively.
 Panels (C,D) show the cumulative distributions for different age groups
 of citation count $c$ and rescaled citation count $R(c)$, respectively.}
\label{fig:one_bis}
\end{figure*}

\begin{table*}[t]
\centering
\caption{The five considered metrics and their bias by age. The difference $\sigma/\sigma_0-1$ quantifies
how much the histogram of the number of top-$1\%$ papers by the metric deviates from the histogram expected
under the hypothesis of ranking not biased by age (see the main text).
The values of $\sigma/\sigma_0-1$ are expressed as multiples of their expected value $\sigma_{dev}=0.11$ 
for a random ranking of the papers (computed as 
explained in the main text).
Values of $\sigma/\sigma_0-1$ smaller than $2\,\sigma_{dev}=0.22$ are reported in bold characters.}
\begin{tabular*}{\hsize}{@{\extracolsep{\fill}}llr}
Metric& Properties &  $\sigma/\sigma_0-1$\cr
\hline
Citation count $c$		& Local metric		 				& $54.64\, \sigma_{dev}$     \cr
PageRank score $p$		& Network-based metric					& $117.36 \, \sigma_{dev}$	\cr
CiteRank $T$			& Network-based metric, time-aware			& $15.91\, \sigma_{dev}$      \cr
Rescaled PageRank $R(p)$	& Network-based metric, time-aware			& $\textbf{1.45} \, \sigma_{dev}$\cr		
Rescaled citation count $R(c)$	& Local metric, time-aware				&$\textbf{0.91} \, \sigma_{dev}$ \cr	
\hline
\end{tabular*}
\label{tab:metrics}
\end{table*}

\subsection{Rescaled PageRank and rescaled citation count}

To compute the rescaled PageRank score $R(p)$ for a given paper $i$, we evaluate 
the paper's PageRank score $p_i$ as well as the mean $\mu_{i}(p)$ and standard deviation $\sigma_{i}(p)$ of 
PageRank score for papers published in a similar time as $i$. 
Time is not measured in days or years, but in number $n$ of
published papers; after labeling the papers in order of decreasing age, $\mu_{i}(p)$ and $\sigma_{i}(p)$ are 
computed over papers $j\in [i-\Delta_{p}/2, i+\Delta_{p}/2]$. The parameter $\Delta_p$ represents the number
of papers in the averaging window of each paper\footnote{In order to have the same number of papers in each averaging window,
a different definition of averaging window is needed for the oldest and the most recent $\Delta_p/2$ papers, for which
we compute $\mu_i$ and $\sigma_i$ over the papers $j\in[1,\Delta_p]$ and $j\in(N-\Delta_p,N]$, respectively.}.
The rescaled score $R_{i}(p)$ of paper $i$ is then computed
as
\begin{equation}
R_{i}(p) = \frac{p_i-\mu_i(p)}{\sigma_i(p)}.
\label{rpr}
\end{equation}
Values of $R(p)$ larger or smaller than zero indicate whether the paper is out- or under-performing, respectively, with respect
to papers of similar age. 
$R_{i}(p)$ represents the $z$-score \cite{kreyszig2010advanced} of paper $i$ within its averaging window. 
For the sake of completeness, we have also tested a simpler rescaled score in the form $R^{(ratio)}_i(p)=p_i/\mu_i(p)$; 
however, $R^{(ratio)}(p)$ fails
to produce a time-balanced ranking due to the fact that $\sigma(p)/\mu(p)$ strongly depends on paper age (see \ref{appendix_alternative}
for details).
In addition, we tested a rescaled score $R^{(year)}(p)$ based on Eq. \eqref{rpr}
where $\mu_i(p)$ and $\sigma_i(p)$ are computed over the papers published in the same 
year as paper $i$. We found that while $R^{(year)}(p)$ is able to suppress large part of PageRank's temporal bias,
its ranking is much less in agreement with the hypothesis of unbiased ranking than the ranking by $R(p)$ 
(see \ref{appendix_alternative} for details). For this reason, we use an
averaging window based on number of publications and not on real time.
This choice is also supported by the findings of refs. \cite{newman2009first} and \cite{parolo2015attention} which suggest that
the role of time in citation networks is better captured by the number of published papers than by real time.

We define the rescaled citation count analogously as 
\begin{equation}
R_{i}(c) = \frac{c_i-\mu_i(c)}{\sigma_i(c)},
\label{rindeg}
\end{equation}
where  $\mu_{i}(c)$ and $\sigma_{i}(c)$ represent the mean and the standard deviation of $c$ computed
over papers $j\in [i-\Delta_{c}/2, i+\Delta_{c}/2]$.
Citation count rescaling was used in refs. \cite{newman2009first, newman2014prediction} to identify
papers that accrue more citations than expected
for papers of similar age under the hypothesis of pure preferential attachment.

The choice of the size of the temporal window deserves some attention: if the size of the temporal window is too large,
one would fall again in a time-biased ranking that is one of the issues that motivate the present
paper. On the other hand, if we choose a too small averaging window, the papers would be only compared
with few other papers and the resulting scores would be too volatile.
Throughout this paper, we set $\Delta_c=\Delta_p=1000$; we refer to \ref{appendix_window}
for further details on the dependence of ranking properties on the averaging window size.
We stress that
the rankings by $R(c)$ and $R(p)$ are only weakly dependent on $\Delta_c$ and $\Delta_p$ (see Fig. \ref{fig:appendix_three}),
and the correlation between the rankings by $R(p)$ obtained with different values
of PageRank's teleportation parameter $\alpha$ is close to one
(Spearman's rank correlation coefficient between the rankings obtained with $\alpha=0.5$ and $\alpha=0.85$
is equal to $0.98$).
These results indicate that the proposed rescaling metrics are robust with respect to variations of their parameters.

\section{Results}
\label{sec:results}

We analyzed the network composed of $L=4,672,812$ citations among $N=449,935$ papers published in APS journals ($1893-2009$).
The dataset was directly provided by the APS following our request at the webpage \url{http://journals.aps.org/datasets},
and was also studied in ref. \cite{medo2011temporal}, among others.

\subsection{Time balance of the rankings}
\label{sec:balance}

Before comparing the performances of the five metrics in recognizing the Milestone Letters (MLs),
we want to determine whether the metrics are biased by age and, if yes, then to which extent.
In agreement with refs. \cite{radicchi2008universality,radicchi2011rescaling}, 
we assume that a fair ranking of scientific papers should be time-balanced in the sense that old and recent 
papers should be equally likely to appear at the top of the ranking by the metric.
Caveats and possible weak points of this assumption are examined in the Discussion section.

To assess the degree of time balance of the five metrics, we perform a statistical test similar to those
proposed in refs. \cite{radicchi2011rescaling, radicchi2012testing}.
We divide the papers into $S=40$ different groups according to their age 
and, for each metric, we compute the number $n_{\alpha}(z)$ of
top-$z\,N$ papers by the metric for each age group $\alpha$, and quantitatively compare the resulting histogram $\{n_{\alpha}(z)\}$ with the 
expected histogram $\{n^{(0)}_{\alpha}(z)\}$ under the hypothesis that the ranking is temporally unbiased.
We set $z=0.01$; results for other small values of $z$ are qualitatively similar.

Fig. \ref{fig:one}A shows that the observed values of $n(0.01)$ for PageRank are far from their expected values under the hypothesis of unbiased
ranking.
For instance, $n_1(0.01)/n_1^{(0)}(0.01)=4.62$ for the age group that contains the oldest $N/40$ papers, 
as opposed to $n_{40}(0.01)=0$ for the age group composed of the most recent $N/40$ papers.
To quantify the degree of time balance of a metric, we compare the standard deviation $\sigma$ of the observed histogram $\{n_{\alpha}(0.01)\}$
with the expected standard deviation $\sigma_0$ under the hypothesis of unbiased ranking.
For a perfectly unbiased ranking, the number $n_{\alpha}^{(0)}$ of nodes from age group $\alpha$
in the top-$z$ by the ranking obeys the multivariate hypergeometric distribution \cite{radicchi2012testing}.
Therefore, we expect on average $n^{(0)}(z)=z\,N/S$ top-$z\,N$ papers for each set, with the standard deviation
\begin{equation}
\label{sigma_0}
 \sigma_0(z)=\sqrt{\frac{z\,N}{S}\,\Biggl(1-\frac{1}{S}\Biggr)\,(1-z)\,\frac{N}{N-1}},
\end{equation}
The observed standard deviation $\sigma(z)$ is computed as
\begin{equation}
 \sigma(z)=\sqrt{\frac{1}{S}\sum_{\alpha=1}^{S}(n_{\alpha}-n_{\alpha}^{(0)})^2}.
 \label{sigma}
\end{equation}
The ratio $\sigma/\sigma_0$ between observed and expected standard deviation quantifies the degree of time balance of the ranking --
we expect this ratio to be close to or lower than (due to fluctuations) one for an unbiased ranking, and significantly larger than one
for a ranking biased by age.
To quantify to which extent the observed values of $\sigma/\sigma_0-1$ are consistent with the hypothesis of unbiased ranking, 
we run a simulation where $0.01\,N$ papers
are randomly assigned to one among $40$ groups, and compute the standard deviation $\sigma_{dev}$ of the observed
deviation $\sigma_{rand}/\sigma_0-1$ according to Eq. \eqref{sigma}. With $10^5$ realizations,
we obtain $\sigma_{dev}=0.11$. 
We always express the observed values of $\sigma/\sigma_0-1$ as multiples of $\sigma_{dev}$ in the following.

We obtain $\sigma/\sigma_0-1=12.91=117.36\,\sigma_{dev}$ for PageRank, which indicates that the ranking is heavily biased.
The heavy bias of PageRank score is also revealed by a comparison of its distribution for nodes from different age groups,
which shows a clear advantage for old nodes (Fig. \ref{fig:one}B).
Fig. \ref{fig:one}C shows that the ranking by the $R(p)$ score is in good agreement with the hypothesis that the ranking is unbiased;
 we find $\sigma/\sigma_0-1=0.16= 1.45\,\sigma_{dev}$.
 The time balance of rescaled PageRank score manifests itself in the collapse of the distributions of the $R(p)$ score for different age groups
 on a unique curve,
 which means that the $R(p)$ score allows us to compare papers of any age on the same scale
 (Fig. \ref{fig:one}D).
In a similar way, the rescaling procedure suppresses the temporal bias of citation count
[$\sigma/\sigma_0-1=0.10=0.91 \,\sigma_{dev}$ for $R(c)$ as compared 
to $\sigma/\sigma_0-1=6.01=54.64\,\sigma_{dev}$ for $c$, see Fig. \ref{fig:one_bis}].
We observe a qualitatively similar suppression of time bias for different choices of the number $S$ of age groups (not shown here).

With respect to the histogram obtained with $R(p)$, 
the histogram $\{n_{\alpha}(0.01)\}$ obtained with the CiteRank algorithm (with the parameters chosen in ref. \cite{maslov2008promise})
presents much larger deviation from the histogram expected under the hypothesis of time-balanced ranking (see Fig. \ref{fig:one}E).
As a result, the value of $\sigma/\sigma_0$ obtained for CiteRank ($\sigma/\sigma_0-1=1.75=15.91\,\sigma_{dev}$ 
with the parameters chosen in ref. \cite{walker2007ranking}) 
is larger than the value obtained for
$R(p)$.
The distributions of CiteRank score $T$ for different age groups do not collapse on a single curve (see Fig. \ref{fig:one}F),
which is directly due to the built-in exponential decay of the teleportation term.
The failure of CiteRank in producing a time-balanced ranking is well exemplified by the behavior of the score distribution for the most recent age group,
whose minimum score (i.e., the smallest score value such that $P(>T)$ deviates from one) is
much larger than for the other distributions, due to a larger teleportation term. 
These findings show that CiteRank score does not allow us to fairly compare papers of different age.

The values of $\sigma/\sigma_0-1$ for the five metrics are summarized in Table \ref{tab:metrics}.

 \begin{figure*}[t]
 \centering
  \includegraphics[scale=0.75,  angle=0]{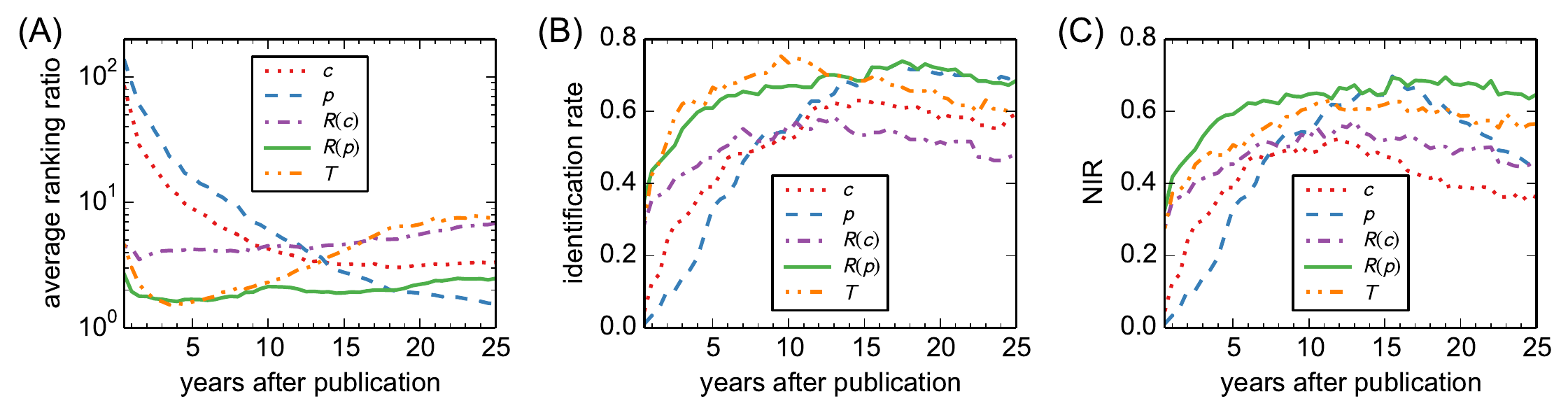}
 \caption{Metrics' performance in ranking the milestone letters as a function of paper age. 
 (A) Dependence of the average ranking ratio $\overline{r}$ on paper age. 
 (B) Dependence of the identification rate $f_{0.01}$ on paper age.
  (C) Dependence of the normalized identification rate $\tilde{f}_{0.01}$ on paper age.
 \label{fig:two}}
\end{figure*}

\begin{figure*}[t]
\centering
 \includegraphics[scale=0.75,angle=0]{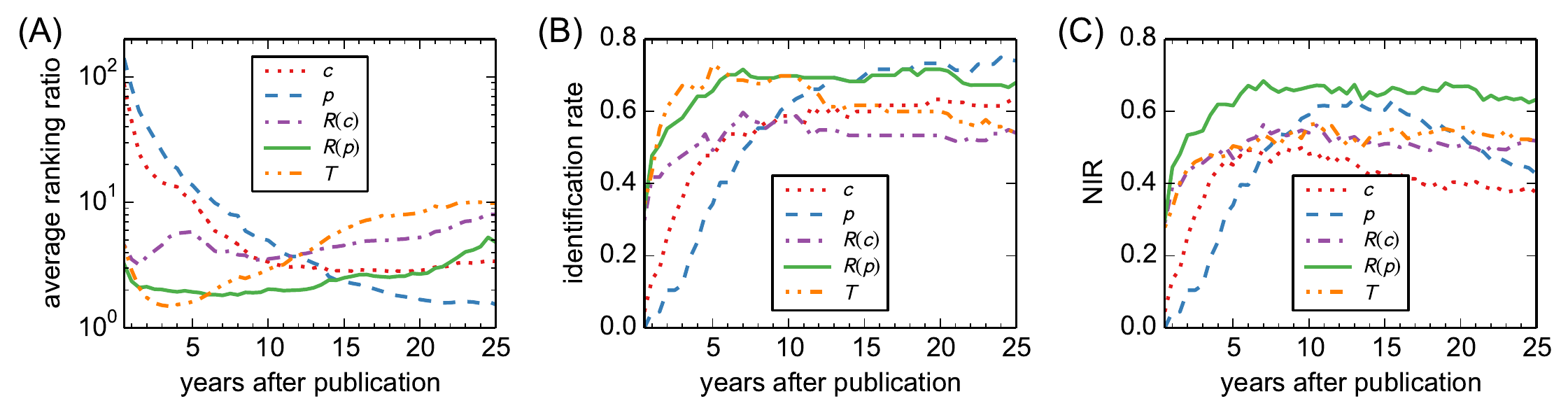}
\caption{Metrics' performance in ranking the APS papers that led to Nobel prize for some of the authors,
listed in the Supplementary Table S1. The figure has been realized with the same procedure used for Fig. \ref{fig:two}. 
 (A) Dependence of the average ranking ratio $\overline{r}$ on paper age.
 (B) Dependence of the identification rate $f_{0.01}$ on paper age.
 (C) Dependence of the normalized identification rate $\tilde{f}_{0.01}$ on paper age.
 We observe a behavior in qualitative agreement with that observed in Fig. \ref{fig:two}.
}
\label{fig:nobel}
\end{figure*}

\begin{table*}[t]
\begin{tabularx}{\linewidth}{|c|c|c|c|X|c|c|l|}
    \caption{Top-15 papers in the APS data as ranked by PageRank score $p$ (asterisks mark the Milestone Letters).  }\\
    \toprule
    \hline
    \textbf{Rank ($p$)}&Rank ($R(p)$)&$p (\times 10^{-5})$&$R(p)$&Title&Year&Journal \\[6pt]
    \midrule
    \endhead
    \hline
$1$& $1$ & $43.32$ & $29.96$ & Self-consistent equations including exchange and correlation effects (W. Kohn, L. Sham)					&1965&\emph{Phys. Rev.}\\
$2$& $36$ & $40.77$ & $24.57$ & Theory of superconductivity (J. Bardeen, L. Cooper, J. Schrieffer)							&1957 &\emph{Phys. Rev.}\\
$3$& $8$ & $35.88$ & $28.58$ & Inhomogeneous electron gas (P. Hohenberg)										&1964 &\emph{Phys. Rev.}\\
$4$& $115$ & $24.74$& $18.64$ & Stochastic problems in physics and astronomy (S. Chandrasekhar) 							&1943 &\emph{Rev. Mod. Phys.}\\
$5$& $137$ & $23.57$ & $17.78$ & The theory of complex spectra (J. Slater)										&1929 &\emph{Phys. Rev.}\\
$6$& $21$ & $23.46$& $26.53$ & $^{*}$A model of leptons (S. Weinberg)								&1967 &\emph{Phys. Rev. Lett.}\\
$7$& $130$ & $22.80$& $18.05$ & Can quantum-mechanical description of physical reality be considered complete? (A. Einstein, B. Podolsky, N. Rosen)	&1935 &\emph{Phys. Rev.}\\
$8$& $140$ & $22.67$& $17.73$ & Crystal statistics. i. A two-dimensional model with an order-disorder transition (L. Onsager)				&1944 &\emph{Phys. Rev.}\\
$9$& $15$ & $22.64$& $27.44$ & Self-interaction correction to density-functional approximations for many-electron systems (J. Perdew)			&1981 &\emph{Phys. Rev. B}\\
$10$& $335$ & $22.39$& $13.17$ & Absence of diffusion in certain random lattices (P. Anderson)								&1958 &\emph{Phys. Rev.}\\
$11$& $16$ & $21.25$& $26.88$ & Scaling theory of localization: absence of quantum diffusion in two dimensions (E. Abrahams)				&1979 &\emph{Phys. Rev. Lett.}\\
$12$& $110$ & $20.67$& $18.83$ & Effects of configuration interaction on intensities and phase shifts (U. Fano)						&1961 &\emph{Phys. Rev.}\\
$13$& $82$ & $19.36 $& $20.86$ & On the constitution of metallic sodium (E. Wigner, F. Seitz)								&1933 &\emph{Phys. Rev.}\\
$14$& $210$ & $18.32$& $15.44$ & On the interaction of electrons in metals (E. Wigner)									&1934 &\emph{Phys. Rev.}\\
$15$& $315$ & $18.25$& $13.53$  & Cohesion in monovalent metals (J. Slater)										&1930 &\emph{Phys. Rev.}\\
\hline
    \bottomrule
\end{tabularx}
\label{table:pr}
\end{table*}

\begin{table*}[t]
\begin{tabularx}{\linewidth}{|c|c|c|c|X|c|c|l|}
    \caption{Top-15 papers in the APS data as ranked by rescaled PageRank score $R(p)$ (asterisks mark the Milestone Letters).  }\\
    \toprule
    \hline
    Rank ($p$)&\textbf{Rank ($R(p)$)}&$p (\times 10^{-5})$&$R(p)$&Title&Year&Journal \\[6pt]
    \midrule
    \endhead
    \hline
$1$  &$1$& $43.32$ & $29.96$ & Self-consistent equations including exchange and correlation effects (W. Kohn, L. Sham)					&1965 &\emph{Phys. Rev.}\\
$63$ &$2$& $11.35$	&$29.63$ & $^{*}$Bose-einstein condensation in a gas of sodium atoms (K. Davis et al.)			&1995 &\emph{Phys. Rev. Lett.}\\
$16$  &$3$& $17.74$ &$29.34 $ & Self-organized criticality: an explanation of the 1/f noise (P. Bak, C. Tang, K. Wiesenfeld)				&1987 &\emph{Phys. Rev. Lett.}\\
$115$  &$4$&$8.60$ &$29.16$ & $^{*}$Large mass hierarchy from a small extra dimension (L. Randall)				&1999 &\emph{Phys. Rev. Lett.}\\
$29$  &$5$& $14.99 $ &$29.01 $ & Pattern formation outside of equilibrium (M. Cross)						&1993 &\emph{Rev. Mod. Phys.}\\
$112$ &$6$& $8.66$ &$28.97$ & Statistical mechanics of complex networks (R. Albert, A.-L. Barabási)				&2002 &\emph{Rev. Mod. Phys.}\\
$181$ &$7$&$7.11$ &$28.95 $ & Review of particle properties (K. Hagiwara et al)						&2002 &\emph{Phys. Rev. D}\\
$3$ &$8$& $35.88$ & $28.58$ & Inhomogeneous electron gas (P. Hohenberg)										&1964 &\emph{Phys. Rev.}\\
$99$ &$9$&$9.35$ &$28.58$ & Evidence of Bose-Einstein condensation in an atomic gas with attractive interactions (C. Bradley et al.)		&1995 &\emph{Phys. Rev. Lett.}\\
$59$ &$10$&$11.65$ &$28.11 $ & Efficient pseudopotentials for plane-wave calculations (N. Troullier, J. Martins)					&1991 &\emph{Phys. Rev. B} \\
$53$ &$11$& $12.11$ &$27.88$ & $^{*}$Teleporting an unknown quantum state via dual classical and Einstein-Podolsky-Rosen channels (C. Bennett et al.)	&1993 &\emph{Phys. Rev. Lett.}\\
$281$ &$12$& $5.99$ &$27.85$ & $^{*}$Negative refraction makes a perfect lens (J. Pendry)					&2000 &\emph{Phys. Rev. Lett.}\\
$216$ &$13$& $6.59$ &$27.59$ & Tev scale superstring and extra dimensions (G. Shiu, S.-H. Tye)								&1998 &\emph{Phys. Rev. D}\\	
$17$ &$14$& $17.54$ &$27.47$ & Diffusion-limited aggregation, a kinetic critical phenomenon (T. Witten)						&1981 &\emph{Phys. Rev. Lett.}\\
$9$ &$15$& $22.64$& $27.44$ & Self-interaction correction to density-functional approximations for many-electron systems (J. Perdew, A. Zunger)	&1981 &\emph{Phys. Rev. B}\\
\hline
    \bottomrule
\end{tabularx}
\label{table:rpr}
\end{table*}

\subsection{Identification of the Milestone Letters}
\label{sec:early_identification}

In the previous section, we have shown that the rankings by the rescaled metrics $R(p)$ and $R(c)$ are consistent with
the hypothesis that the ranking is not biased by paper age.
While different works have recently emphasized the importance of removing the bias by age of citation-performance metrics
for a fair ranking of scientific publications \cite{radicchi2008universality, radicchi2011rescaling}
and researchers \cite{kaur2013universality},
the possible positive effects of time-balanced rankings with respect to biased rankings remain largely unexplored.

Ref. \cite{chen2007finding} analyzed the APS dataset and found that PageRank is able to recognize 
old papers that are universally important for physics. 
They also noted that PageRank is based on a diffusion process that drifts towards old papers
(see ref. \cite{mariani2015ranking} for a general analysis of this aspect) and, as a consequence, it inevitably favors old papers.
Since the rescaling procedure that we propose solves this issue, it is thus plausible to conjecture that with respect
to the PageRank algorithm, rescaled PageRank allows us to identify seminal papers earlier.

In this section, we use the APS dataset and the 
list of Milestone Letters (MLs) chosen by editors of Physical Review Letters (see Supplementary Table S1 for the list of MLs. Supplementary 
material for this manuscript is available at \url{http://www.sciencedirect.com/science/article/pii/S1751157716301729}).
to address the two following research questions:
\begin{enumerate}
 \item \emph{Is there a significant gap between the performance of rescaled PageRank and PageRank in
 identifying the MLs short after publication? If there is a substantial gap, does it close down after a certain number of years after
 publication?}
 \item \emph{Do network-based indicators outperform indicators based on simple citation count in recognizing the MLs?}
\end{enumerate}

To compare the ranking positions of the MLs by the five different metrics, 
the ranking of Milestone Letter $i$ is computed $t$ years after its publication. We calculate the
ratio of $i$'s ranking position $r_i(s,t)$ by metric $s$ and $i$'s best ranking
position $\min_{s}\{r_{i}(s,t)\}$ among all considered metrics. To characterize
the overall performance of metric $s$ in ranking the MLs, we average the ranking
ratio over $i$ and obtain $\overline{r}(s,t)$ (see \ref{sec:computation_details} for 
computation details). 
The resulting quantity is
referred to as the \emph{average ranking ratio} of metric $s$ for the Milestone Letters $t$ years after their publication.
A good metric is expected to have as low $\overline{r}(\cdot,t)$ as possible --
the minimum value $\overline{r}(\cdot,t)=1$ is only achieved by a metric that always outperforms the others
in ranking the milestone papers of age $t$. 
Note that the average ranking ratio reduces to average ranking position if we do not normalize the ranking position $r_i(s,t)$ by
$\min_{s}\{r_{i}(s,t)\}$.
However, the average ranking position of the target papers by a certain metric is 
extremely sensitive to the ranking positions of the least-cited target papers,
as opposed to the robustness of the average ranking ratio
with respect to removal of the least-cited papers from the set of target papers 
(see \ref{appendix_average} for details). This property motivates the use of ranking ratio to compare the ranking
positions of the MLs by the different metrics.

The dependence of $\overline{r}(s,t)$ on paper age $t$ measured in years after publication is shown in Fig. \ref{fig:two}A.
Due to the suppression of time bias, rescaled PageRank score $R(p)$ has a large advantage with respect to the original PageRank score $p$ 
for papers of small age.
Since the PageRank algorithm is biased towards old nodes, 
the performance gap between $R(p)$ and $p$ gradually decreases with age and vanishes $18$ years after publication.
By contrast, the CiteRank algorithm exponentially penalizes older nodes and, as a consequence,
the performance gap between $R(p)$ and $T$ is minimal for recent papers, and CiteRank score $T$ can even outperform
$R(p)$ during the first six years after publication.
When paper age becomes sufficiently larger than CiteRank's temporal timescale ($\tau=2.6$ years here, as chosen 
in refs. \cite{walker2007ranking, maslov2008promise}),
older papers are strongly penalized by the CiteRank's teleportation term and, as a result, CiteRank is markedly outperformed by rescaled PageRank.
The same behavior is observed also for other values of CiteRank time-decay parameter $\tau$ (see \ref{appendix_citerank}).
The local metrics $c$ and $R(c)$ are outperformed by $R(p)$ in ranking the MLs of every age, which indicates that network centrality
brings a substantial advantage in ranking highly significant papers with respect to simple and rescaled citation count.

While the average ranking ratio $\overline{r}$ takes into account all the MLs, 
it is also interesting to measure the age-dependence of the identification rates of the metrics, defined as
the fraction $f_{x}(t)$ of MLs that were ranked among the top $x\,N$ papers by the 
metric when they were $t$ years old \footnote{The identification rate is related to \emph{recall}, a standard measure in the literature of
recommendation systems \cite{lu2012recommender}.} (see Fig. \ref{fig:two}B).
Rescaled PageRank $R(p)$ and CiteRank score $T$ markedly outperform the other metrics in identifying the milestone papers
in the first years after publication. 
The performance gap between $R(p)$ and the citation-based indicators $c$ and $R(c)$ remains significant during the whole observation lapse.
Analogously to what we observed for the average ranking ratio,
the performance gap between $R(p)$ and $p$ gradually decreases with paper age
and vanishes $15$ years after publication, which is similar to the crossing point at $18$ years after publication 
observed for the average ranking ratio.
CiteRank has a small advantage with respect to rescaled PageRank in the first years after publication,
whereas for older papers CiteRank's identification rate drops to the value achieved by simple citation count $c$.

It is worth to observe that
the temporal bias of a certain metric affects the behavior of both $\overline{r}(t)$ and $f_{0.01}(t)$ for that metric:
as we observe in \ref{sec:whole_dataset},
a metric biased towards old (like PageRank) or recent papers naturally performs better in
identifying old or recent MLs, respectively.
One natural way to understand this effect is to consider a normalized identification rate $\tilde{f}_{0.01}(t)$
(hereafter abbreviated as NIR),
such that the contribution of each identified ML $i$ of age $t$ (i.e., a ML ranked in the 
top $0.01\,N$ of the ranking) to $\tilde{f}_{0.01}(t)$ 
is smaller than one if the metric favors papers that belong to the same age group as paper $i$ (see \ref{sec:computation_details} for the 
mathematical definition).
In other words, when evaluating the performance of a given metric, the normalized identification rate $\tilde{f}_{0.01}(t)$ 
takes into account both the temporal balance and the identification power of the metric.
The behavior of $\tilde{f}_{0.01}(t)$ for the five metrics is shown in Fig. \ref{fig:two}C.
After an initial increasing trend for all the metrics, the normalized identification rate of both $p$ and $c$ decline due to their temporal bias;
by contrast, the same quantity remains relatively stable for both $R(p)$ and $R(c)$.
According to $\tilde{f}_{0.01}(t)$, rescaled PageRank outperforms CiteRank for papers of every age. 
This is due to the fact that the ranking by CiteRank is not unbiased and,
as a consequence, CiteRank's performance is often penalized by the NIR for small age $t$ due to the algorithm's bias towards recent nodes.

Our analysis assumes that a ML should be ranked as high as possible by
a good metric for scientific significance.
On the other hand, many outstanding contributions to physics are not included in the list of MLs. 
To show that our results also hold for an alternative choice of groundbreaking papers,
we consider a list of $67$ APS papers that led to Nobel Prize for some of the authors (see Supplementary Table S1 for the list
of papers).
The results for this list of benchmark papers are shown in Fig. \ref{fig:nobel} and are qualitatively similar to those shown in Fig. \ref{fig:two},
which indicates that our findings are robust with respect to modifications of the benchmark papers' list.

While Fig. \ref{fig:two} concerns the metrics' performance averaged over the whole set of MLs,
the Supplementary Movie available 
at our webpage [\url{http://www.ddp.fmph.uniba.sk/~medo/physics/RPR/}] 
shows the simultaneous dynamics of the ranking positions by $p$ and $R(p)$ of all individual 
MLs for the first $15$ years after 
publication\footnote{Accordingly, only the $73$ MLs that are at least $15$ years old 
at the end of the dataset are included in the movie.}. 
The movie shows that rescaled score $R(p)$ has a clear advantage with respect to PageRank score $p$ in the first years after publication
for most of the MLs.
As the MLs become sufficiently old, their position in the plane gradually tends to converge to the diagonal where the ranking position 
by $p$ is equal to the ranking position by $R(p)$, which is in agreement with
the crossing between PageRank's and rescaled PageRank's performance curves observed in Figs. \ref{fig:two}A-\ref{fig:two}B.

In principle, one might consider a comparison of the final ranking positions (i.e., the ranking positions computed on the whole dataset) 
of the target papers by a certain metric \cite{dunaiski2012comparing, dunaiski2016evaluating} 
instead of the age-dependent evaluation of the metrics
introduced above.
But this kind of comparison would miss our key point -- the strong dependence of metrics' performance on paper age.
In addition, the strong dependence of metrics' performance on paper age shown in this section makes the outcome of such evaluation
strongly dependent on the age distribution of the target papers we aim to identify.
This issue is discussed in \ref{sec:whole_dataset} and potentially concerns any performance evaluation carried out on
a fixed snapshot of the network.
By contrast, the outcomes presented in this paragraph (how well do the different metrics perform as a function of paper age)
are little sensitive to the exact age distribution of the target papers.

\subsection{Top papers by PageRank and rescaled PageRank}

To get an intuitive understanding of the properties of PageRank and its rescaled version, it is 
instructive to look at the top-$15$ papers according to $p$ and $R(p)$ computed on the whole dataset,
reported in Table II and Table III, respectively.
Although only one ML appears in the top $15$ by $p$ (ranked $6$th, see Table II), among the non-MLs there are papers of exceptional significance, 
such as the letter that proposed the popular Einstein-Podolsky-Rosen experiment (ranked $7$th); the paper that introduced a 
fundamental tool in many-body systems, 
Slater's determinant (ranked $5$th);
the paper that presented the famous exact solution of the two-dimensional Ising model (ranked $8$th).
This confirms that PageRank is highly effective 
in finding relatively old papers of outstanding significance -- referred to as ``scientific gems'' in ref. \cite{chen2007finding} --
which has led to the interpretation of
PageRank score as a ``lifetime achievement award`` for a paper \cite{maslov2008promise}.
Nevertheless, the most recent paper in Table II is from $1981$ -- $28$ years old with respect to the dataset's ending point in $2009$.

In the top-$15$ by $R(p)$, we find both old papers (the oldest is from $1964$, $45$ years old in $2009$)
and recent papers (the most recent is from $2002$, $7$ years old in $2009$).
Four out of $15$ top-papers are MLs, which is an additional confirmation of the quality of the ranking by $R(p)$.
We emphasize that while both PageRank and rescaled PageRank feature prominent papers in their top-$15$, 
the detailed performance analysis described in the previous section is essential in order to fully understand the behavior of the two metrics.


\section{Discussion}
\label{sec:discussion}

Motivated by the recent publication of the list of Milestone Letters by the Physical Review Letters editors,
we performed an extensive cross-evaluation of different data-driven metrics of scientific impact of research papers
with respect to their ability to identify papers of exceptional significance.
We studied the network of citations between papers in the Physical Review corpus, which is recognized to be a
comprehensive proxy for scientific research in physics \cite{redner2005citation, radicchi2009diffusion, radicchi2011rescaling}.
The main assumption of our analysis is that although not all the most important papers in the Physical Review corpus 
are covered by the Milestone Letters list, a good paper-level metric is expected to rank the Milestone Letters as high as possible
due to their outstanding significance.
We find a clear performance gap between network-based metrics ($p,R(p),T$) and local metrics based only on the number
of received citations ($c,R(c)$).
This finding suggests that the use of citation counts to rank scientific papers is
sub-optimal; additional research will be needed to assess whether network-based article-level metrics can be used
to construct author-level metrics more effective than
the currently used metrics -- such as the popular $h$-index \cite{hirsch2005index} -- that are only based on citation counts
and neglect network centrality.

We have shown that the proposed rescaled PageRank $R(p)$ suppresses PageRank's well-known bias
against recent papers much better than the CiteRank algorithm does.
As a result,
the proposed rescaled PageRank $R(p)$ provides a superior performance than PageRank and CiteRank in ranking recent
and old milestone papers, respectively.
There are still two possible ranking errors---false positives and false negatives---that 
have not been addressed in this manuscript. Young papers at the top of the ranking by the rescaled PageRank may be false 
positives because the citation spurt that they have experienced may stop which will eventually force
them out of the ranking's top as well as out from the group of possibly highly significant papers. 
By contrast, the so-called ''sleeping beauties`` that receive a large part of citations long after
they are published \cite{ke2015defining} are likely to be under-evaluated by the rescaled PageRank. 
Assessing the extent to which false positives and false negatives affect the ranking by rescaled PageRank, 
and by other relevant metrics as well, goes beyond the scope of our paper yet it constitutes a much needed step in future research.
The analysis of larger datasets which include papers from diverse fields is another natural next step for future research.
As different academic disciplines adopt different citation practices \cite{bornmann2008citation},
the rescaling procedure proposed in this paper may need to be extended to also
remove possible ranking biases by academic field.

The assumptions behind our definition of time balance and the computation of the rescaled scores deserve attention as well.
In agreement with refs. \cite{radicchi2008universality, radicchi2011rescaling},
the definition of time balance of a ranking adopted in this article
requires that the likelihood that a paper is ranked at the top by a time-balanced metric is independent of paper age.
Our definition of ranking time balance is implicitly based on the 
assumption that the number of highly significant
papers grows linearly with system size.
While this assumption seems reasonable for the Physical Review corpus whose journals apply strict acceptance criteria for submitted papers,
it might need to be reconsidered when analyzing
larger datasets which include recently emerging high-acceptance journals (both mega-journals~\cite{bjork2015have} 
and predatory journals~\cite{xia2015publishes}).
In other words, the exponential growth of the number of published papers \cite{redner2005citation,wang2013quantifying, sinatra2015century}
does not necessarily correspond to an exponential growth of the number of highly significant papers.
The issue is delicate (see ref. \cite{sarewitz2016pressure} for a recent insight) and 
will need to be addressed in future research on bibliometric indicators.



An important general question remains open: which inherent properties of a network
determine if PageRank-like methods will outperform local metrics or not?
We conjecture that in citation networks, the observed success of network-based metrics in identifying highly significant papers
might be related to the tendency of high-impact papers to cite other high-impact papers \cite{bornmann2010scientific}.
Despite recent efforts \cite{mariani2015ranking, medo2015identification, ghoshal2011ranking, fortunato2008approximating},
which network properties make the PageRank algorithm succeed or fail remains a largely unexplored problem which we will further
investigate in future research.

Our work constitutes a particular instance of a general methodology -- the comparison of the outcomes of quantitative variables with 
a ground-truth established by experts -- which can be applied for metric evaluation in several kinds of systems, such
as movies \cite{spitz2014measuring,wasserman2015cross} or the network of scientific authors \cite{radicchi2009diffusion}.
In the domain of research evaluation, this methodology is particularly relevant since bibliometric indices are increasingly used in practice
-- often uncritically and in questionable ways \cite{wilsdon2015we, hicks2015bibliometrics} -- and scholars from diverse field have
produced a plethora of possible impact metrics \cite{van2010metrics}, especially those aimed at
assessing researchers' productivity and impact. 
Motivated by the results obtained in this article, we encourage the creation of 
lists of groundbreaking papers also for other scientific domains,
which can lead to a richer understanding and more accurate benchmarking of quantitative metrics for scientific significance.
Our findings constitute a benchmark for article-level metrics of scientific significance,
and can be used as a baseline to assess the performance of new indicators in future research.

From a practical point of view, 
improving the effectiveness of paper impact metrics has the potential to improve not only the current bibliometric practices, but also
our ability to discover relevant papers in 
online platforms that collect academic papers and use automated methods to sort them.
In this respect, our 
findings suggest that rescaled PageRank can be used as an operational tool to identify the most significant
papers on a given topic.
Suppose that a researcher enters a new research 
field and wants to study the most important works in that field.
If we provide him/her with the top papers as ranked by PageRank, the researcher will only know the oldest papers and will not be informed about recent
lines of research.
On the other hand, by providing him/her with the top papers as ranked by rescaled PageRank, he/she will know both
old significant papers and recent works that have attracted considerable attention, leading to a more complete overview of the field.
To allow researchers to experience the benefits of a time-balanced ranking method,
we developed an interactive Web platform which is available at the address \url{http://www.sciencenow.info}.
In this platform, users can browse the rankings of the APS papers by $R(p)$ year by year,
investigate the historical evolution of each paper's ranking position by $R(p)$,
and check the ranking positions and the scores of each researcher's publications.

\section{Conclusions}
\label{sec:conclusions}

We presented a detailed analysis of the performance of different quantitative metrics with respect to
their ability to identify the Milestone Letters selected by the Physical Review Letters editors.
Our findings indicate that: (1) a direct rescaling of citation count and PageRank score is an effective way
to suppress the temporal bias of these two metrics; (2) rescaled PageRank $R(p)$ is the best-performing metric overall,
as it outperforms PageRank and CiteRank in identifying recent and old milestone papers, respectively, 
and it outperform citation-based indicators for papers of every age.
The presented results indicate that the combination of network centrality and time 
holds promise for improving some of the tools currently used to rank scientific publications, which could bring valuable benefits for
quantitative research assessment and design of Web academic platforms.

\section*{Acknowledgements}

We wish to thank Giulio Cimini, Matthieu Cristelli, Luciano Pietronero, Zhuo-Ming Ren, Andrea Tacchella, Giacomo Vaccario, 
Alexandre Vidmer and Andrea Zaccaria
for inspiring discussions and useful suggestions.
This work was supported by the EU FET-Open Grant No. 611272 (project Growthcom). 
The authors declare that they have no competing financial interests.
Correspondence and requests for materials should be addressed to M. S. Mariani~(email: manuel.mariani@unifr.ch).

\bibliographystyle{elsarticle-num}

\newpage 

\appendix

\onecolumngrid

\addcontentsline{toc}{section}{Appendices}

\addtocontents{toc}{\protect\setcounter{tocdepth}{-1}}

\section{Average ranking position vs. average ranking ratio}
\label{appendix_average}

We show here that the average ranking position of the MLs is extremely sensitive to the ranking position of the least-cited MLs,
whereas the average ranking ratio is stable with respect to removal of the least-cited MLs.
For simplicity, in this Appendix we consider the rankings computed on the whole dataset. 
In formulas, the average ranking position $\overline{r_{raw}}(s)$ of the MLs by metric $s$ is defined as
\begin{equation}
 \overline{r_{raw}}(s)=\frac{1}{M}\sum_{i\in\mathcal{M}}r_i(s),
\end{equation}
where $r_i(s)$ denotes the ranking position of paper $i$ by metric $s$ normalized by the total number of papers: $r_i=1/N$ and $r_i=1$
correspond to the best and the worst paper in the ranking, respectively.

In section \ref{sec:early_identification}, we mention that little-cited 
papers can bias the average ranking position of the target 
papers by a certain metric.
To illustrate this point, consider first the following ideal example.
Consider two target papers $A$ and $B$. Paper $A$ is ranked $10$th by metric $M_1$ and 
$1,000$th by metric $M_2$, whereas paper $B$ is ranked $20,000$ by metric $M_1$ and $15,000$ by metric $M_2$.
The average ranking position for the set of papers $\{A,B\}$ is equal to $10,005$ and to $8,000$ for metric $M_1$ and $M_2$, respectively.
This means that according to average ranking position, 
metric $M_2$ outperforms metric $M_1$, despite having not been able to place any of the two papers in the top-$100$.

A qualitatively similar situation occurs also in the APS dataset, as the following example shows.
The milestone letter ''Element No. 102`` [Phys. Rev. Lett. 1.1 (1958): 18] is cited only five times within the APS data.
Its ranking position by $R(p)$ ($r(R(p))=0.22$) is thus much larger than the MLs' average ranking position $\overline{r_{raw}}(R(p))=0.016$ by
$R(p)$.
Only few MLs are little cited -- for instance, only four out of $87$ MLs are not among the top-$10\%$ papers by citation count.
To which extent do these little-cited papers affect $\overline{r_{raw}}$ for the different metrics? 
By denoting with $\overline{r'}_{raw}(R(p))$ the average computed on the subset of $83$ MLs which does not include the four least-cited MLs, 
we obtain $\overline{r'}_{raw}(R(p))=0.009$,
which is smaller than $\overline{r}_{raw}(R(p))=0.016$ by a factor around $1.8$.
The effect is even larger for citation count: we have $\overline{r'}_{raw}(c)=0.009$ against the original
value $\overline{r}_{raw}(c)=0.020$ -- the ratio between the two averages is larger than two.

By using the average ranking ratio, we only compare the ranking within the chosen set of metrics \emph{for each individual paper} and,
as a consequence, the average is stable with respect to removal of the least-cited MLs. 
This can be illustrated by again excluding the four least-cited MLs
from the computation of $\overline{r}(R(p))$,
and by comparing the corresponding values $\overline{r'}(R(p))$ of the average ranking ratio with the values computed over all the MLs.
Among the five metrics, the largest variation is observed for PageRank, for which
$\overline{r'}(p)/\overline{r}(p)=1.03$
-- i.e., the removal of the least-cited MLs has only a small effect on the average ranking ratios for the five metrics.

\section{Assessing the metrics' performance on the whole dataset}
\label{sec:whole_dataset}

 \begin{figure*}[t]
 \centering
  \includegraphics[scale=0.75,  angle=0]{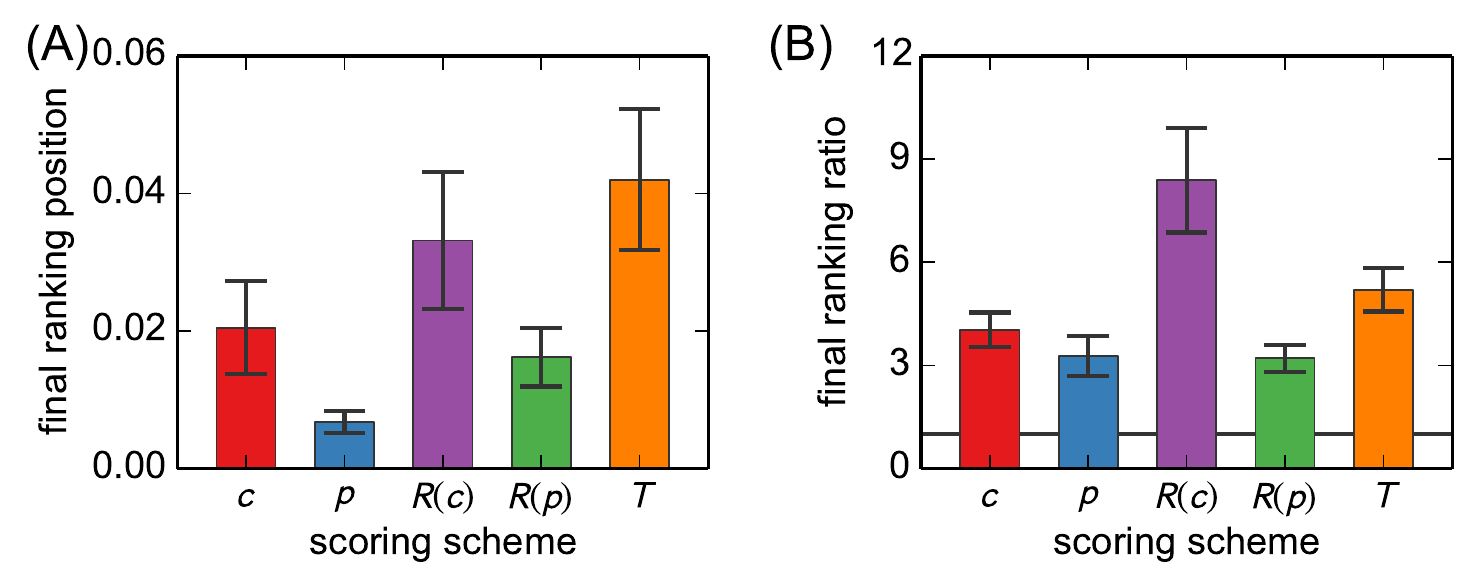}
 \caption{Values of the average ranking position $\overline{r_{raw}}$  (panel A)
 and of the average ranking ratio $\overline{r}$ (panel B) of the MLs 
 for the five metrics
 computed on the whole dataset (1893-2009); the error bars represent the standard error of the mean.
 \label{fig:histograms}}
\end{figure*}

\begin{figure}[t]
\centering
\includegraphics[scale=1,angle=0]{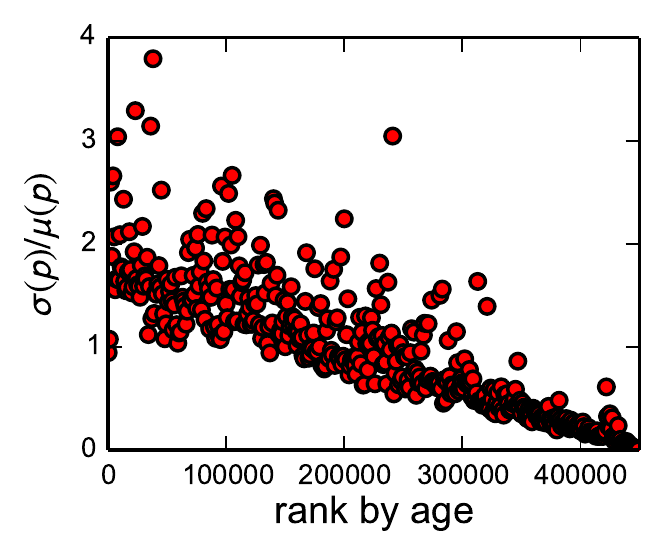}
\caption{Dependence of $\sigma(p)/\mu(p)$ on paper age; the values of $\mu(p)$ and $\sigma(p)$ are calculated over the papers' averaging windows.}
\label{fig:mu_and_sigma}
\end{figure}

\begin{figure}[t]
\centering
 \includegraphics[scale=0.9,angle=0]{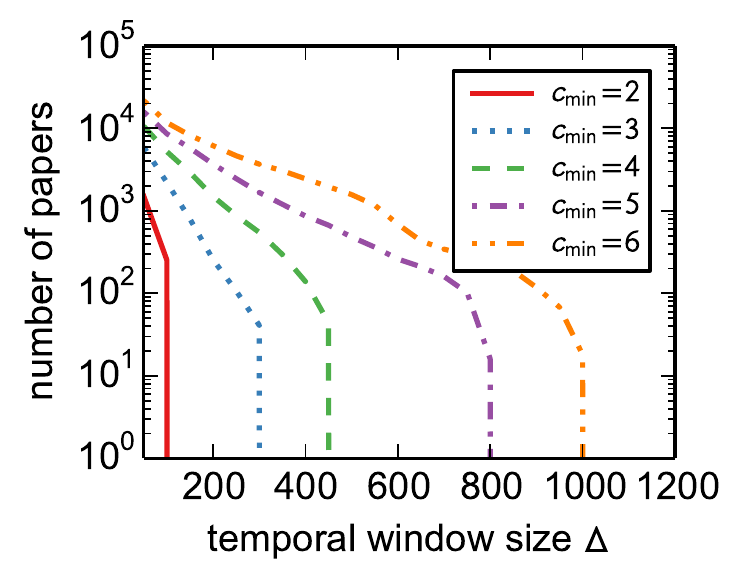}
\caption{Number of papers
whose averaging window contains less than five papers that received at least $c_{min}$ citations as a function of $\Delta$.
For $\Delta \geq 1000$, each paper is compared with at least five papers cited at least five times.}
\label{fig:appendix_one}
\end{figure}

\begin{figure*}[t]
\centering
 \includegraphics[scale=0.9,angle=0]{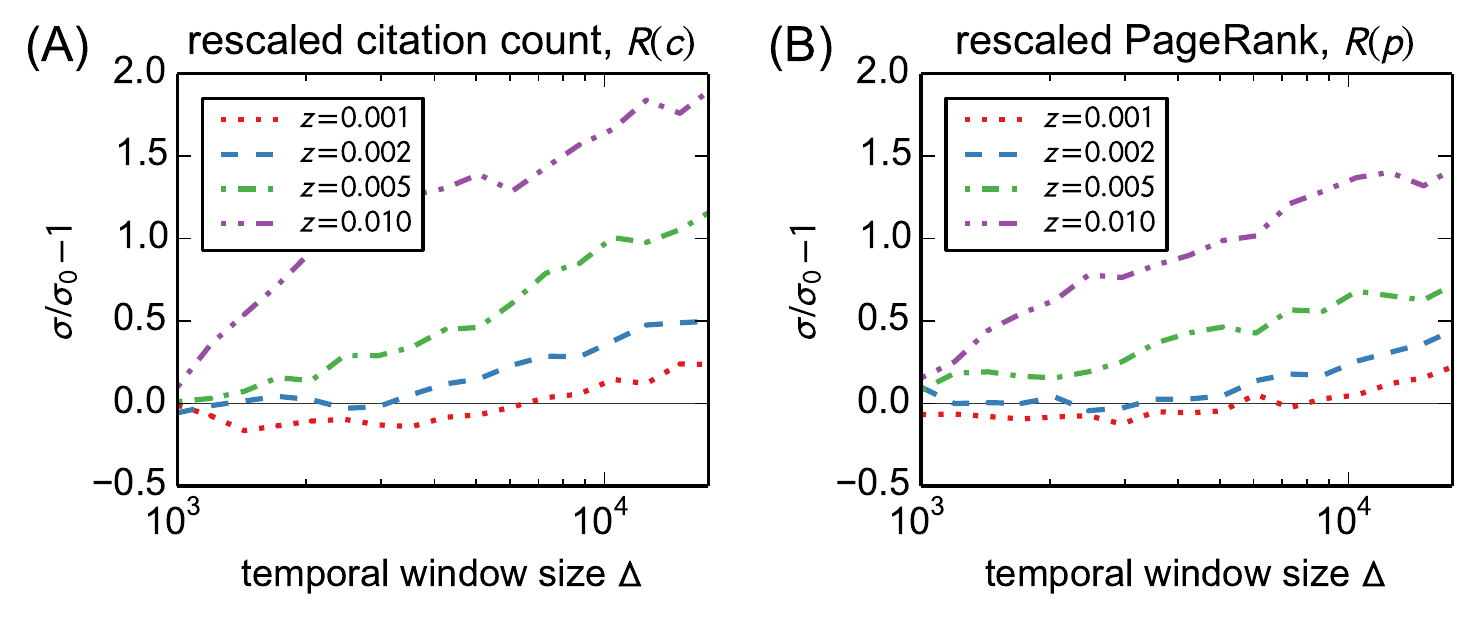}
\caption{\emph{Left panel:}
Deviation $\sigma/\sigma_0-1$ for the ranking by rescaled citation count $R(c)$ 
as a function of $\Delta_c$ for different values of $z$.
\emph{Right panel:} Deviation $\sigma/\sigma_0-1$ for the ranking by rescaled PageRank score $R(p)$ 
as a function of $\Delta_p$ for different values of $z$. The horizontal black line marks the expected value $\sigma/\sigma_0-1=0$
for an unbiased ranking. }
\label{fig:appendix_two}
\end{figure*}

\begin{figure}[t]
\centering
 \includegraphics[scale=0.9,angle=0]{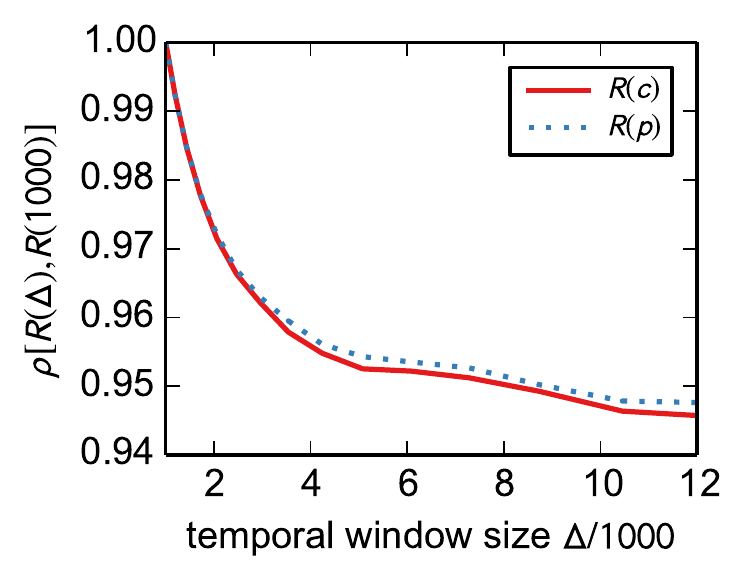}
\caption{Spearman's ranking correlation between the 
rescaled score $R(\Delta)$ and the rescaled score $R(\Delta=1000)$ used in the main text.}
\label{fig:appendix_three}
\end{figure}

Fig. \ref{fig:histograms}A shows the values of the average ranking position $\overline{r_{raw}}(s)$ for the five metrics
computed on the whole dataset:
according to $\overline{r_{raw}}(s)$, PageRank and rescaled PageRank outperform the other metrics.

While the average ranking position of the MLs is a simple quantity to evaluate the metrics, some MLs are
relatively little cited and, as a result,
their low ranking position can strongly bias the average ranking position.
We refer to \ref{appendix_average} for a detailed discussion of this issue.
To solve this problem, we defined the ranking ratio in the main text.
Fig. \ref{fig:histograms}B shows the measured values of the average ranking ratio $\overline{r}$ based on the rankings computed on the whole dataset.
This simple measure would suggest that $R(p)$ and, to a lesser extent, $p$ and $c$ outperform $R(c)$ and CiteRank.
Given the small gap between $p$ and $R(p)$, one might be tempted to conclude that the rescaling procedure does not bring substantial benefits
in the identification of significant papers.
However, the rank analysis presented in Fig. \ref{fig:histograms} includes the contribution
of both old and recent MLs, whereas a close inspection reveals that the metrics perform in a drastically different way depending on the age of the
target papers, as shown in Figure 3 and discussed in Section \ref{sec:early_identification}.

This point can be also illustrated by using the rankings computed on the whole dataset. 
To show this, we divide the $87$ MLs into three equally-sized groups of MLs according to their age.
By considering only the oldest $M/3=29$ MLs as target papers, we obtain $\overline{r}(p)=1.1$ whereas $\overline{r}(R(p))=5.5$.
By contrast, by considering only the $M/3$ most recent MLs as target papers, we obtain $\overline{r}(p)=7.3$ whereas $\overline{r}(R(p))=1.7$.
While this result shows a clear advantage of PageRank and rescaled PageRank for the oldest and for the most recent MLs, respectively,
there exists a fundamental difference between
the performance gaps observed for the oldest and the most recent MLs.
The bias of PageRank towards old nodes (Fig. \ref{fig:one}A) makes it indeed easier for the metric to 
find old significant papers.
On the other hand, rescaled PageRank does not benefit from any bias in ranking the most recent MLs as the ranking by the metric
is not biased by paper age (Fig. \ref{fig:one}C).
It is thus crucial to realize that when we compute the rankings on the whole dataset,
the value of the average ranking ratio by the metrics depends on the age distribution
of the important papers that we aim to identify. Were we using the rankings computed on the whole dataset 
for evaluation and were we only considering the oldest (most recent) 
$29$ MLs as target papers, we would have concluded that PageRank (rescaled PageRank) is by far the best-performing metric.
These observations demonstrate that an evaluation of the metrics based on the whole dataset is strongly biased by the
age distribution of the target items and, for this reason, unreliable as a tool to assess metrics' performance.

\section{Alternative rescaling equations}
\label{appendix_alternative}

Eq. \eqref{rpr}
forces the rescaled score $R_{i}(p)$ of a paper $i$ to have mean value equal to zero and standard deviation equal to one,
independently of its age (i.e., independently of $i$). Fig. 2C shows that this rescaling is sufficient to achieve a time-balanced
ranking of the papers.
We consider now 
a simple rescaling in the form $R^{(ratio)}_i(p):=p_i / \mu_i(p)$.
While the mean value of this score is equal to one, one can show that its standard deviation is given by 
\begin{equation}
 \sigma\bigr[R^{(ratio)}_i(p)\bigl]=\sqrt{\mathrm{E}_i\bigr[(R^{(ratio)}_i(p))^2\bigl]-\mathrm{E}_i\bigr[R^{(ratio)}_i(p)\bigl]^2}
 =\sqrt{\frac{\mathrm{E}_i[p_{i}^{2}]}{\mu_i(p)^2}-1}
 =\frac{\sigma_{i}(p)}{\mu_{i}(p)},
\end{equation}
where $E_i[\cdot]$ denotes the expectation value within the averaging window of paper $i$.
Fig. B.6 shows that $\sigma(p)/\mu(p)$ strongly depends on node age in the APS dataset.
As a result, the ranking by $R^{(ratio)}(p)$ is strong biased towards old nodes ($\sigma/\sigma_0-1=79.81\,\sigma_{dev}$).

We also considered a variant of our method where the rescaled scores are still computed with Eq. \eqref{rpr},
but $\mu_i(p)$ and $\sigma_i(p)$ are computed over the papers published
in the same year as paper $i$. The resulting rescaled score $R^{(year)}(p)$ produces a ranking that is much less in agreement with the hypothesis of
unbiased ranking ($\sigma/\sigma_0-1=15.55\,\sigma_{dev}$) than the ranking by $R(p)$. For this reason, the definition of papers' averaging window
adopted in the main text is based on number of publications and not on real time.
However, $R^{(year)}(p)$ is still preferable to the original scores when the aim is to compare papers
of different age.
Also note that $R^{(year)}(p)$ might be preferable if one is interested in a ranking of the papers where
each publication year is represented by the same number of papers, apart from statistical fluctuations.

\section{Dependence of the properties of the rankings by $R(c)$ and $R(p)$ on the temporal window size $\Delta$}
\label{appendix_window}

As described in the main text, the rescaled scores $R_{i}(c)$ and $R_{i}(p)$ of a certain paper $i$ are obtained by comparing its score 
with the scores of the nodes that belong to its ``averaging windows`` $j\in[i-\Delta_c/2,i+\Delta_c/2]$ and $j\in[i-\Delta_p/2,i+\Delta_p/2]$, respectively.
To motivate the choice $\Delta_p=\Delta_c=1000$ adopted in the main text,
we start by observing that
the size of the averaging window should be neither too large nor too small.
A large window would include papers of significantly different age, which would turn out to be ineffective in removing
the temporal biases of the metrics -- note that the ranking by $R(p)$ is perfectly correlated with the 
ranking by $p$ for $\Delta_p= N$.
On the other hand, we want $\Delta_c$ and $\Delta_p$ to be sufficiently large to avoid that some papers are only compared with little-cited papers,
which is likely to happen for a small window due to the skewed shape of the citation count distribution \cite{medo2011temporal}.

To understand the possible drawbacks of a too small averaging window, we compute the number $N(c_{min})$ of papers
whose averaging windows contain less than five papers that received at least $c_{min}$ citations.
The results are shown in Fig. \ref{fig:appendix_one}.
For $\Delta\leq800$, the averaging windows of a nonzero number of papers have less than five papers with at least five received citations.
We restrict our choice to the range $\Delta\geq 1000$, for which no paper's average window has less than five papers cited at least five times.

To evaluate the ability of the rescaling procedure to suppress the bias of the metrics,
we estimate the deviation $\sigma/\sigma_0-1$ of the standard deviation ratio $\sigma/\sigma_0$ from the expected value (one) for an unbiased ranking
(see the main text for details).
Fig. \ref{fig:appendix_two} reports the behavior of the deviation $\sigma/\sigma_{0}-1$
as a function of $\Delta_p$ and $\Delta_c$ for 
different selectivity values $z$.
The upward trends of Fig. \ref{fig:appendix_two} suggest that in order to reduce the ratio $\sigma/\sigma_0$,
it is convenient to choose $\Delta_p$ and $\Delta_c$ as small as possible.
Hence, the choice $\Delta_c=\Delta_p=1000$ allows us to obtain an histogram close to the expected 
unbiased histogram -- $\sigma/\sigma_0$ values are
close to one for all the values of $z$ represented in the figure -- and, at the same time, to avoid that some nodes are only compared with little cited
nodes, as discussed above for \ref{fig:appendix_one}.

An important observation is that the correlations between the rankings obtained with different values
of $\Delta$ and the ranking obtained with $\Delta=1000$ are close to one (Fig. \ref{fig:appendix_three}),
which means that the rescaling procedure is robust against variation of the averaging window sizes $\Delta_c$ and $\Delta_p$.

\section{Dependence of CiteRank performance on its parameter $\tau$}
\label{appendix_citerank}

\begin{figure*}[t]
\centering
 \includegraphics[scale=0.9,angle=0]{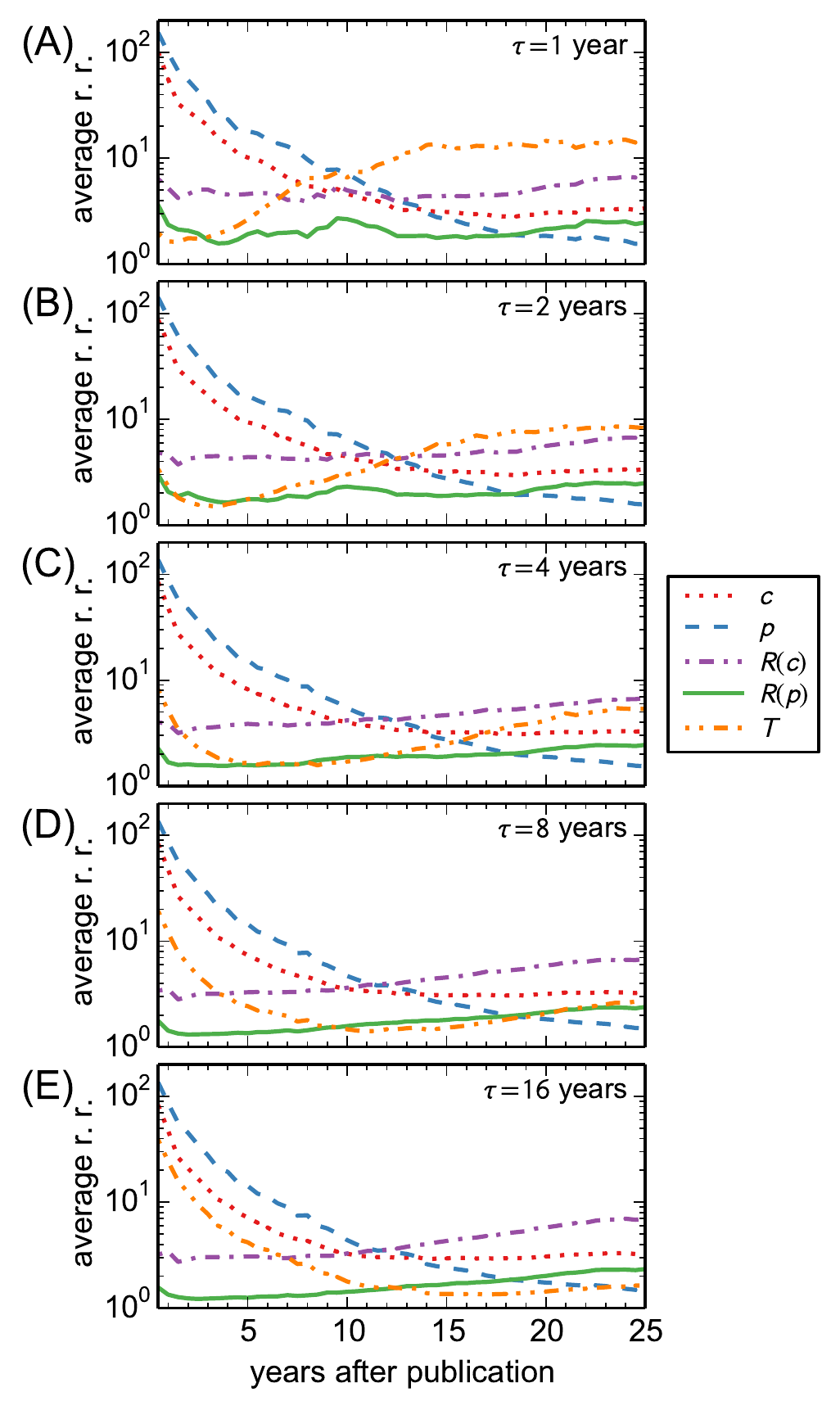}
\caption{Dependence of the average ranking ratio $\overline{r}$ on paper age, for five different values of
CiteRank parameter $\tau$.
}
\label{fig:citerank}
\end{figure*}

Fig. \ref{fig:citerank} shows the dependence of the average ranking ratio $\overline{r}$ on paper age, for five different values of
CiteRank parameter $\tau$.
The figure shows that the behavior of CiteRank's performance strongly depends on the choice of its parameter.
When the parameter is small (panel A, $\tau=1$ year), CiteRank performance is optimal (lowest average ranking ratio) 
for very recent papers, and gradually worsens with paper age.
As $\tau$ increases (moving from panel A to E), the minimum point of CiteRank's average ranking ratio gradually shifts toward older nodes.
When $\tau$ is sufficiently large (panel E, $\tau=16$ years), CiteRank behavior is qualitatively similar to that of PageRank,
and its performance gradually improves with paper age -- this is indeed consistent with the fact that $T\to p$ in the limit $\tau\to\infty$.

\section{Dependence of ranking ratio and identification rate on paper age}
\label{sec:computation_details}

To assess the ranking of each Milestone Letter $t$ years after its publication,
we compute the rankings each $\Delta t=183$ days (results for different choices
of $\Delta t$ are qualitatively similar).
At each computation time $t^{(c)}$, only the $N(t^{(c)})$ papers (with their links) published before time $t^{(c)}$ are considered for the
scores' and rankings' computation, and each ML contributes to the ranking ratio $\overline{r}(s,t)$
corresponding to its age $t$ at time $t^{(c)}$. 
This procedure allows us to save computational time with respect to computing the rankings of each ML \emph{exactly} $t$ years after its publication,
because it requires fewer ranking computations.

In formulas, the average ranking ratio $\overline{r}(s,t=k\,\Delta t)$ for $t$-years old papers is defined as
\begin{equation}
 \overline{r}(s,t=k\,\Delta t)=\frac{1}{M(t)}\sum_{t^{(c)}}\sum_{i\in\mathcal{M}} \delta\Bigl(\floor{(t^{(c)}-t_{i})/\Delta t},k\Bigr) 
 \times \frac{r(s,i;t^{(c)})}{\min_{s'}\{r(s',i;t^{(c)})\}},
 \end{equation}
where we used $k=0.5,1,1.5,2,\dots$ for Fig. \ref{fig:two}B;
in the equation above, $r(s,i;t^{(c)})$ 
denotes the ranking position of ML $i$ at time $t^{(c)}$ according to metric $s$,
$M(t)$ denotes the number of MLs that are at least $t$ years old at the end of the dataset, $\floor{x}$ denotes
the largest integer smaller than or equal to $x$, $\delta(x,y)$ denotes the Kronecker delta function of $x$ and $y$.
Hence, at each computation time $t^{(c)}$, each ML $i$ published before time $t^{(c)}$
gives a contribution $\hat{r}(s,i;t^{(c)})$ to the average ranking ratio $\overline{r}(s,t=k\,\Delta t)$
for papers of age $t^{(c)}-t_{i}$.
Similarly, the identification rate $f_{x}(t)$ is computed as
\begin{equation}
 f_x(s,k\,\Delta t)=\frac{1}{M(t)}\sum_{t^{(c)}}\sum_{i\in\mathcal{M}} \delta\Bigl(\floor{(t^{(c)}-t_{i})/\Delta t},k\Bigr)\,
 \times \chi(r(s,i;t^{(c)})\leq x),
 \end{equation}
where $\chi(r(s,i;t^{(c)})\leq x)$ is equal to one if paper $i$ is among the top $x\,N(t^{(c)})$ papers in the ranking by metric $s$ at time $t^{(c)}$,
equal to zero otherwise.

To define the normalized identification rate (NIR) of a metric, at each computation time $t^{(c)}$ we divide the $N(t^{(c)})$ papers
into $40$ groups according to their age, analogously to what we did in section \ref{sec:balance} to evaluate
the temporal balance of the metrics.
The NIR of metric $s$ is then defined as
\begin{equation}
 \tilde{f}_x(s,k\,\Delta t)=\frac{1}{M(t)}\sum_{t^{(c)}}\sum_{i\in\mathcal{M}} \delta\Bigl(\floor{(t^{(c)}-t_{i})/\Delta t},k\Bigr)\,
 \times \chi(r(s,i;t^{(c)})\leq x)\,y(n(s,i;t^{(c)})),
\label{nir}
 \end{equation}
 where $y(n(s,i;t^{(c)}))$ is a decreasing function of the fraction $n(s,i;t^{(c)})$ of 
 nodes that belong to the same age group of node $i$ and are ranked among the top $x\,N(t^{(c)})$
by metric $s$.
Denoting by $n_{0}(i;t^{(c)})=1/40$ the expected value of $n(\cdot,i;t^{(c)})$ for an unbiased ranking,
we set $y(n(s,i;t^{(c)}))=(n(s,i;t^{(c)})/n_{0}(i;t^{(c)}))^{-1}$ if $n(s,i;t^{(c)})>n_{0}(i;t^{(c)})$ (i.e., if the metric tends to favor papers that belong
to the same age group as paper $i$), whereas $y(n(s,i;t^{(c)}))=1$ if $n(s,i;t^{(c)})\leq n_{0}(i;t^{(c)})$.
According to Eq. \eqref{nir}, if the
identified ML belongs to an age group which is over-represented in 
top $x\,N(t^{(c)})$ by the factor of four, it only counts as $1/4$ in the normalized identification rate.

\end{document}